\documentclass{pramana}

\usepackage{graphicx,amsmath,bm}
\usepackage{epstopdf}
\usepackage{microtype}
\usepackage{xcolor}
\begin{document}

\title{State Feedback Control and Observer Based Adaptive Synchronization of 
\\ Chaos in a Memristive Murali-Lakshmanan-Chua Circuit}


\author{A. ISHAQ AHAMED\textsuperscript{1} \and M. LAKSHMANAN\textsuperscript{2,*}}
\affilOne{\textsuperscript{1} Department of Physics, Jamal Mohamed College, (Affiliated to Bharathidasan University), Tiruchirappalli-620020, India \\}
\affilTwo{\textsuperscript{2} Department of Nonlinear Dynamics, School of Physics, Bharathidasan University,Tiruchirappalli-620024, India}


\twocolumn[{

\maketitle

\corres{lakshman.cnld@gmail.com}

\msinfo{23 January 2020}{23 January 2020}{23 January 2020}

\begin{abstract}
In this paper we report the control and synchronization of chaos in a Memristive Murali-Lakshmanan-Chua circuit. This circuit,  introduced by the present authors in 2013, is basically a non-smooth system having two discontinuity boundaries by virtue of it having a flux controlled active memristor as its nonlinear element. While the control of chaos has been effected using state feedback techniques, the concept of adaptive synchronization and observer based approaches have been used to effect synchronization of chaos. Both of these techniques are based on state space representation theory which is well known in the field of control engineering. As in our earlier works on this circuit, we have derived the Poincar\'{e} Discontinuity Mapping (PDM) and Zero Time Discontinuity Mapping (ZDM) corrections, both of which are essential for realizing the true dynamics of non-smooth systems. Further we have constructed the  observer and controller based canonical forms of the state space representations, have set up the Luenberger observer, derived the controller gain vector to implement state feedback control and  calculated the gain matrices for switch feed back and finally performed parameter estimation for effecting observer based adaptive synchronization. Our results obtained by numerical simulation include time plots, phase portraits, estimation of the parameters and convergence of errors graphs and phase plots showing complete synchronization.

\end{abstract}



}]


\doinum{12.3456/s78910-011-012-3}
\artcitid{\#\#\#\#}
\volnum{123}
\year{2016}
\pgrange{23--25}
\setcounter{page}{23}
\lp{25}

\section{Introduction}
Chaotic systems are characterised by their high sensitivity to even infinitesimal changes in their initial conditions. As a result these systems, by their intrinsic nature, defy attempts at control or synchronization. Nevertheless many techniques have been proposed by a large group of researchers to control and synchronize chaotic systems. Control of chaos refers to a process wherein a judiciously chosen perturbation is applied to a chaotic system, in order to realize a desirable behaviour \cite{gonzalez04}. Since the seminal contribution by Ott, Grebogi and Yorke in 1990 \cite{ogy90} the concept of control of chaos has been modified and developed by many researchers \cite{rajasekar93} and applied to a large number of physical systems \cite{boc00}. Synchronization of chaos, on the other hand, can be described as a process wherein two or more chaotic systems (either equivalent or non-equivalent) adjust a given property of their motion to a common behaviour, due to coupling or forcing. This may range from complete agreement of trajectories to locking of phases \cite{ml96,ml03,boc02}.

In this paper we describe the general principles of control of chaos using state feedback mechanism and synchronization of chaotic systems using observer based adaptive techniques. Further using these, we report the control of chaos in a single Memristive Murali-Lakshmanan-Chua (MLC) oscillator and the synchronization of chaos in a two coupled Memristive MLC  oscillator system. The paper is organized as follows. In Sec. 2 we give a brief introduction of the Memristive MLC circuit, its circuit realization, its circuit equations and their normalized forms and the description of the circuit as a non-smooth system. In Sec. 3 the various algorithms for the control of chaos are outlined. In Sec. 4 the control of chaos in the Memristive MLC circuit using state feed back control technique is dealt with. Similarly in Secs. 5 and 6 the concept of synchronization of chaos and its realization are explained, while in Sec. 7 the observer based adaptive synchronization of chaos in a system of two coupled Memristive MLC oscillator is described. Finally in Sec. 8 the results and further discussions are given.  
\section{Memristive Murali-Lakshmanan-Chua Circuit} 
\begin{figure}[!t]
\centering{
\includegraphics[width=.8\columnwidth]{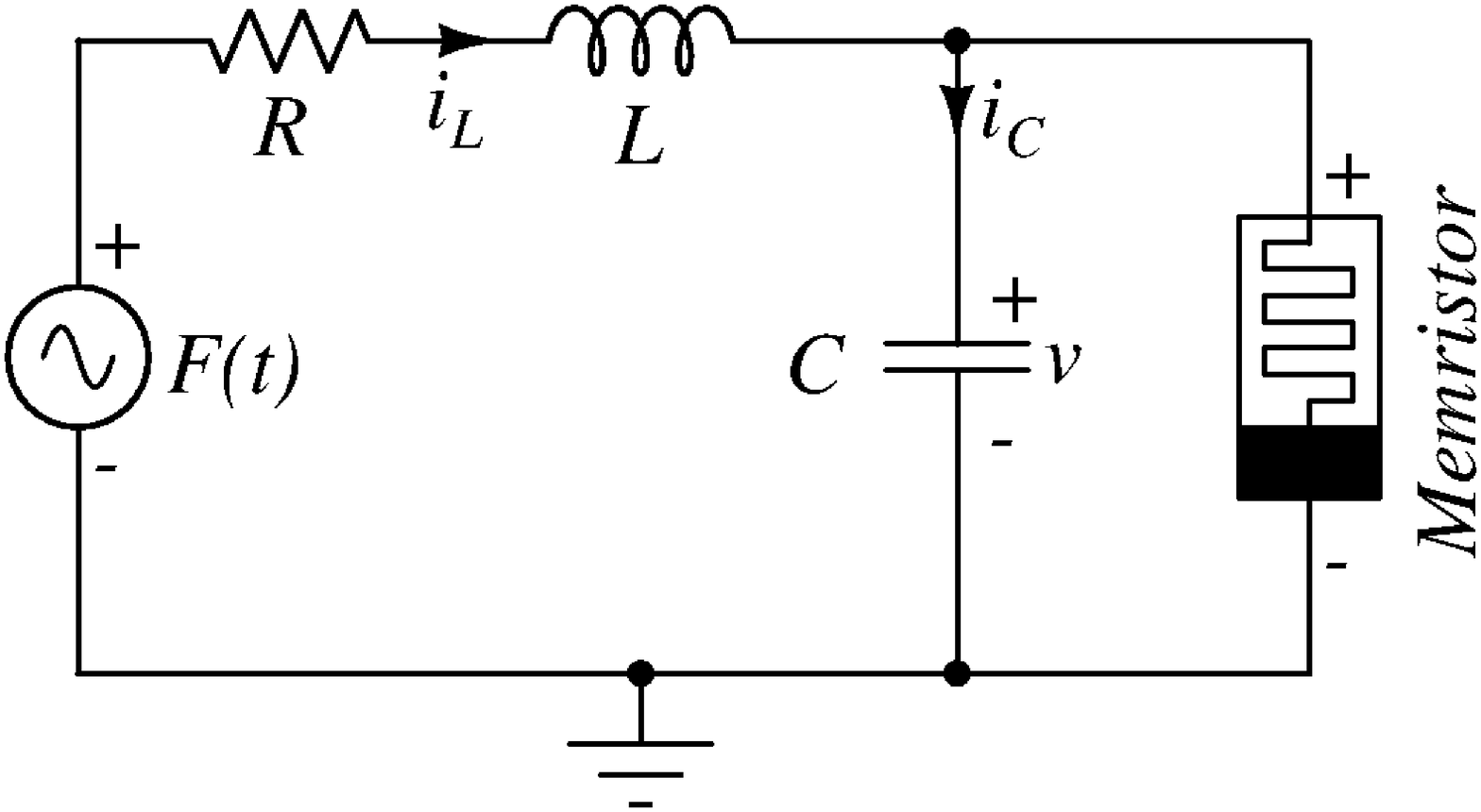}}
\caption{The memristive MLC circuit}
\label{fig:mmlc_cir}
\end{figure}

\begin{figure*}
	\centering{
\includegraphics[width=1.6\columnwidth]{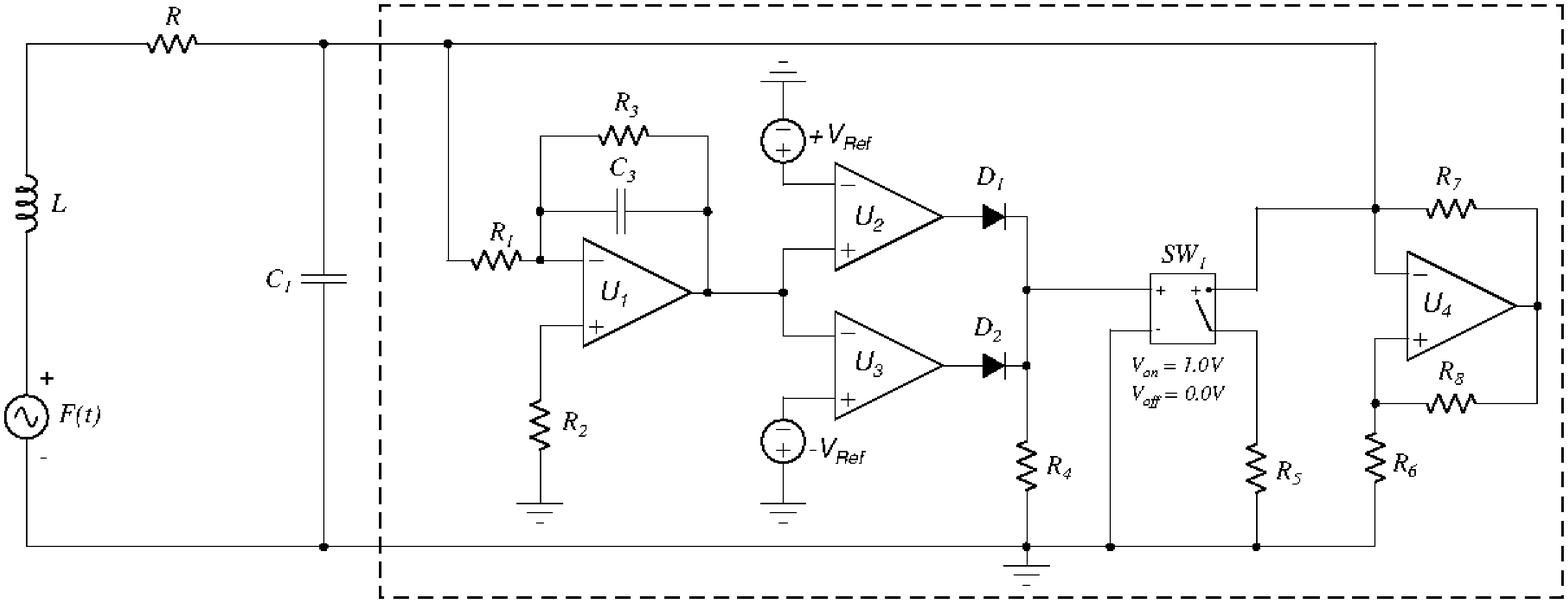}}		\caption[Multisim Prototype of the Memristive MLC Circuit] {A Multisim Prototype Model of a memristive MLC circuit. The memristor part is shown by the dashed outline. The parameter values of the circuit  are fixed as $ L = 21 mH$, $R = 900 \Omega$, $C_1 = 10.5nF$. The frequency of the external sinusoidal forcing is fixed as $\nu_{ext} = 8.288 kHz$ and the amplitude is fixed as $F = 770 mV_{pp} $ ( peak-to-peak voltage). }
		\label{fig:prototype}
\end{figure*}

The memristive MLC circuit was introduced by the present authors \cite{icha13} by replacing the Chua's diode in the classical Murali-Lakshmanan-Chua circuit with an active flux controlled memristor as its non-linear element. The analog model of the memristor used in this work was desinged by \cite{icha11}. The schematic of the memristive MLC circuit is shown in Fig. \ref{fig:mmlc_cir}, while the actual analog realization based on the prototype model for the memristor is shown in Fig. \ref{fig:prototype}.


Applying Kirchoff's laws, the circuit equations can be written as a set of autonomous ordinary differential equations (ODEs) for the flux $\phi(t)$, voltage $v(t)$, current $i(t)$ and the time $p$ in the extended coordinate system as
\begin{eqnarray}
\frac{d\phi}{dt}  & = & v, \nonumber \\
C\frac{dv}{dt}  & = & i - W(\phi)v,  \nonumber \\ 
L \frac{di}{dt}  & = &  -v -Ri +F\sin( \Omega p),\nonumber \\
\frac{dp}{dt}  & = & 1.
	\label{eqn:mlc_cir}
\end{eqnarray}
Here $W(\phi)$ is the memductance of the memristor and is as defined in \cite{itoh08},
\begin{equation}
W(\phi) = \frac{dq(\phi)}{d\phi} = \left\{
					\begin{array}{ll}
					G_{a_1}, ~~~ | \phi  | > 1  \\
					G_{a_2}, ~~~ | \phi  | \leq 1,	
					\end{array}
				\right.
	\label{eqn:W}
\end{equation}
where $G_{a_1}$ and $G_{a_2}$ are the slopes of the outer and inner segments of the characteristic curve of the memristor respectively. We can rewrite Eqs. (\ref{eqn:mlc_cir}) in the normalized form as
\begin{eqnarray}
\dot{x}_1  & = & x_2, \nonumber \\
\dot{x}_2  & = & x_3-W(x_1)x_2, \nonumber 
\end{eqnarray}
\begin{eqnarray}
\dot{x}_3  & = & -\beta(x_2+x_3) + f \sin(\omega x_4),\nonumber \\
\dot{x}_4  & = &  1.
\label{eqn:mlc_nor}
\end{eqnarray}
Here dot stands for differentiation with respect to the normalized time $\tau$ (see below) and $W(x_1)$ is the normalized value of the memductance of the memristor, given as
\begin{equation}
W(x_1) = \frac{dq(x_1)}{dx_1} = \left\{
		\begin{array}{ll}
		a_1, ~~~ | x_1  | > 1 \\
		a_2, ~~~ | x_1  | \leq  1
		\end{array}
	\right.
	\label{eqn:W_nor}
\end{equation}
where $ a_1 = G_{a_1}/G $ and $a_2 = G_{a_2}/G$  are the normalized values of $G_{a_1} $ and $G_{a_2} $  mentioned earlier and are negative. The rescaling parameters used for the normalization are
\begin{eqnarray}
x_1 = \frac{G\phi}{C},x_2 = v, x_3 = \frac{i}{G}, x_4 = \frac{Gp}{C}, G = \frac{1}{R}, \\ \nonumber 
\beta = \frac{C}{LG^2}, \omega = \frac{\Omega C}{G} = \frac{2\pi \nu C}{G},\tau = \frac{Gt}{C},f = F\beta.
	\label{eqn:rescale} 
\end{eqnarray}
In our earlier work on this memristive MLC circuit, see \cite{icha13}, we reported that the addition of the memristor as the nonlinear element converts the system into a piecewise-smooth continuous flow having two discontinuous boundaries, admitting \emph{grazing bifurcations}, a type of discontinuity induced bifurcation (DIB). These grazing bifurcations were identified as the cause for the occurrence of hyperchaos, hyperchaotic beats and transient hyperchaos in this memristive MLC system. Further we have reported \emph{discontinuity induced Hopf and Neimark-Sacker bifurcations} in the same circuit, refer \cite{icha17}. Thus the memristive  MLC circuit shows rich dynamics by virtue of it being a non-smooth system. Hence we give a brief description of the memristive MLC circuit in the frame work of non-smooth bifurcation theory.

\subsection{Memristive MLC Circuit as a Non-smooth System}

The memristive MLC circuit is a piecewise-smooth continuous system by virtue of the discontinuous nature of its nonlinearity, namely the memristor. An active flux controlled memristor is known to switch state with respect to time from a more conductive ON state to a less conductive OFF state and vice versa at some fixed values of flux across it, see \cite{icha17}. In the normalized coordinates this switching is found to occur at $x_1 = +1$ and at $x_1 = -1$. These switching states of the memristor give rise to two discontinuity boundaries or switching manifolds, $\Sigma_{1,2}$ and $\Sigma_{2,3}$ which are symmetric about the origin and are defined by the zero sets of the smooth functions $H_i(\mathbf{x},\mu) = C^T\mathbf{x}$, where $C^T = [1,0,0,0]$ and $\mathbf{x}= [x_1,x_2,x_3,x_4]$, for $i=1,2$. Hence $H_1(\mathbf{x}, \mu) = (x_1-x_1^\ast)$, $x_1^\ast = -1$ and $H_2(\mathbf{x}, \mu) = (x_1-x_1^\ast)$, $x_1^\ast = +1$, respectively. Consequently the phase space $\mathcal{D}$ can be divided into three subspaces $S_1$, $S_2$ and $S_3$ due to the presence of the two switching manifolds. The memristive MLC circuit can now be rewritten as a set of smooth ODEs 
\begin{footnotesize}
\begin{equation}
\dot{x}(t) = 
	\left\lbrace 
			\begin{array}{l}
		F_{1,3}(\mathbf{x},\mu), \, H_1(\mathbf{x}, \mu)< 0 \,\& \, H_2(\mathbf{x},\mu)> 0, \,\mathbf{x} \in S_{1,3}\\ 
		\\
		F_2(\mathbf{x},\mu), \quad H_1(\mathbf{x}, \mu) >0 \, \& \, H_2(\mathbf{x}, \mu) < 0,\mathbf{x} \in S_2  
			\end{array}
		\right.
	\label{eqn:smooth_odes}
\end{equation}
\end{footnotesize}
where $\mu$ denotes the parameter dependence of the vector fields and the scalar functions. The vector fields $F_i$'s are
\begin{equation}
 F_i(\mathbf{x},\mu) =  \left (	\begin{array}{c}
					x_2			\\
				-a_i x_2+x_3	\\
				-\beta x_2 -\beta x_3 +f sin(\omega x_4)	\\	
				1 
				\end{array}
		\right ), \mathrm{i\;=\;1,2,3}
\label{eqn:vect_field}
\end{equation}
where we have $a_1 = a_3$. 

The discontinuity boundaries $\Sigma_{1,2}$ and $\Sigma_{2,3}$ are not uniformly discontinuous. This means that the degree of smoothness of the system in some domain $\mathcal{D}$ of the boundary is not the same for all points $x \in \Sigma_{ij}\cap \mathcal{D}$. This causes the memristive MLC circuit to behave as a non-smooth system having a degree of smoothness of either {\textit{one}} or {\textit{two}}. In such a case it will behave either as a \emph{Filippov system} or as a \emph{piecewise-smooth continuous flow} respectively, refer Appendix A in \cite{icha17}.

\subsection{Equilibrium Points and their Stability}
In the absence of the driving force, that is if $f=0$, the memristive MLC circuit can be considered as a three-dimensional autonomous system with vector fields given by
\begin{equation}
 F_i(\mathbf{x},\mu) =  \left (	
 					\begin{array}{c}
							x_2			 	\\
						-a_i x_2+x_3  		\\
						-\beta x_2 -\beta x_3	\\	
					\end{array}
		\right ), \mathrm{i\;=\;1,2,3}.
\label{eqn:vect_field3d}
\end{equation}
This three dimensional autonomous system has a trivial equilibrium point $E_0$, two {\textit{admissible equilibrium}} points $E_{\pm}$ and two {\textit{boundary equilibrium}} points $E_{B\pm}$.\\The trivial equilibrium point is given as
\begin{equation}
E_0 = \{(x_1,x_2,x_3)|x_1=x_2=x_3=0\}
\end{equation}
The two admissible equilibria $E_{\pm}$ are
\begin{equation}
E_{\pm} = \{(x_1,x_2,x_3)|x_2=x_3=0,x_1^*= \textrm{constant and not equal to } \pm 1 \}
\end{equation}

\begin{figure}[!t]
	\centering
	\resizebox{\columnwidth}{!}
		{\includegraphics{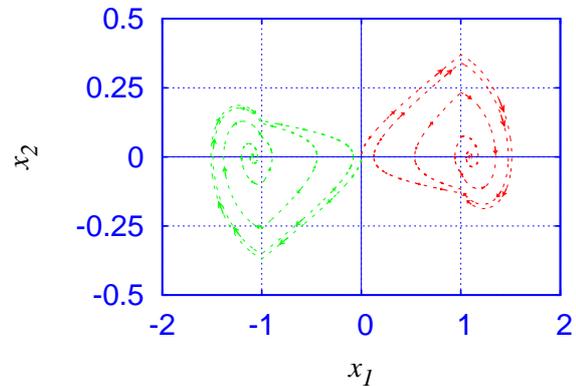}}		
	\caption[Equilibrium Points]{Figure showing the equilibrium points $E_{\pm}$ in the subspaces $S_1$ and $S_3$ for the parameter value above $\beta_c = 0.8250$. The initial conditions are  $x_1 = 0.0$, $x_2 = 0.01$, $x_3 = 0.01$ for the fixed point $E_{+}$ in the subspace $S_3$ and 
 	$x_1 = 0.0$, $x_2 = -0.01$, $x_3 = -0.01$ for the fixed point $E_{-}$ in the subspace $S_1$.}
 		\label{fig:mmlc_limitcycle}
\end{figure} 
The two boundary equilibrium points are 
\begin{equation}
E_{B\pm} = \{(x_1,x_2,x_3)|x_2=x_3=0,\hat{x}_1= \pm 1 \}
\end{equation}
The multiplicity of equilibrium points arises because of the non-smooth nature of the nonlinear function, namely $W(x_1)$ given in Eq. (\ref{eqn:W_nor}). To find the stability of these equilibrium states, we construct the Jacobian matrices $N_i, \, i = 1,2,3$ and evaluate their eigenvalues at these points,
\begin{equation}
N_i =   \left ( \begin{array}{ccc}
				0      & 1    	&  0  		 \\
				0      &-a_i    &  1 		 \\
				0      &-\beta  &  -\beta 	\\				 
				\end{array}
		\right),  \text{i\;=\;1,2,3}.
	\label{eqn:Jac}
\end{equation}
The characteristic equation associated with the system $N_i$ in these equilibrium states is
\begin{equation}
\lambda^3 + p_2\lambda^2 + p_1 \lambda = 0,
	\label{eqn:chac}
\end{equation}
where {\it{$\lambda$}}'s are the eigenvalues that characterize the equilibrium states and $\it{p_i}$'s are the coefficients, given as $p_1 = \beta(1+a_i)$ and $p_2 = (\beta + a_i)$. The eigenvalues are 
\begin{equation}
\lambda_1 = 0, \,\lambda_{2,3} = \frac{-(\beta+a_i)}{2} \pm \frac{\sqrt{(\beta - a_i)^2-4 \beta}}{2}.
	\label{eqn:eigen}
\end{equation} 
where $ i = 1,2,3$. Depending on the eigenvalues, the nature of the equilibrium states differ. 
\begin{enumerate}

\item
When $(\beta - a_i)^2 = 4 \beta$, the equilibrium state will be a stable/unstable star depending on whether $(\beta+a_i)$ is positive or not.

\item
When $(\beta - a_i)^2 > 4 \beta$, the equilibrium state will be a saddle.

\item
When $(\beta - a_i)^2 < 4 \beta$, the equilibrium state will be a stable/unstable focus.
\end{enumerate}
For the third case, the circuit admits self oscillations with natural frequency varying in the range \\$\sqrt{\left [(\beta - a_1)^2-4 \beta \right ]} / 2 < \omega_o < \sqrt{\left [(\beta - a_2)^2-4 \beta \right ]}/2$. \\It is at this range of frequency that the memristor switching also occurs. 

As the vector fields $F_1(\mathbf{x},\mu)$ and $F_3(\mathbf{x},\mu)$ are symmetric about the origin, that is $F_1(\mathbf{x},\mu) = F_3(-\mathbf{x},\mu)$, the admissible equilibria $E_{\pm}$ are also placed symmetric about the origin in the subspaces $S_1$ and $S_3$. These are shown in Fig. \ref{fig:mmlc_limitcycle} for a certain choice of parametric values.

\begin{figure}[!t]
	\centering
	\resizebox{\columnwidth}{!}
		{\includegraphics{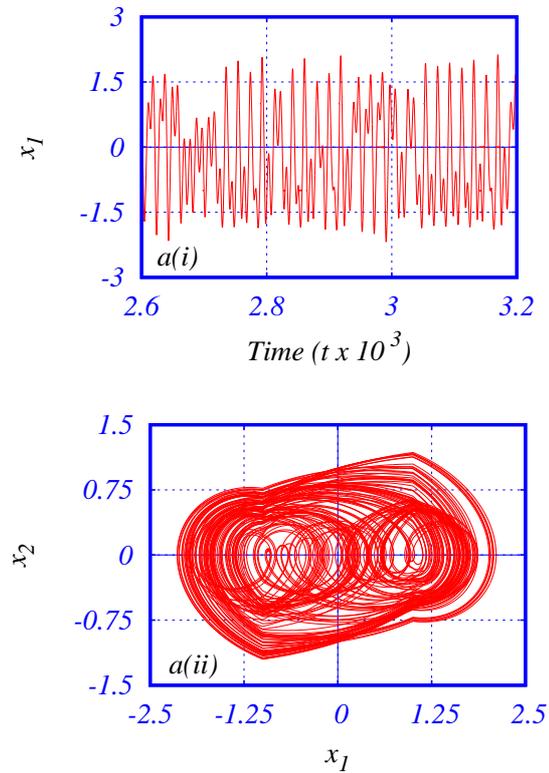}}
		\caption[Dynamics of the Memristive MLC Circuit before application of state feedback control] {The chaotic dynamics of the memristive MLC oscillator arising due to sliding bifurcations occurring in the circuit, with a(i) the time plot of the $x_1$ variable and a(ii) phase portrait in the $(x_1-x_2)$ plane. The step size is assumed as $h = \frac{1}{1000}(2\pi/\omega)$, with $\omega = 0.65$ and $f = 0.20$.}
	\label{fig:control1}
\end{figure}

\subsection{Sliding Bifurcations and Chaos}
Let us assume the bifurcation points at the two switching manifolds to be 
\begin{equation}
E_{B\pm} = \{(x_1,x_2,x_3)|x_2 \neq 0, x_3=0,\hat{x}_1= \pm 1. \}
	\label{eqn:slid_points}
\end{equation}
Then we find from Eqs. (\ref{eqn:vect_field3d}) that $F_2(x,\mu) \neq F_1(x,\mu)$ at $x \in \Sigma_{1,2}$ and $F_2(x,\mu) \neq F_3(x,\mu)$ at $x \in \Sigma_{2,3}$. Under such conditions the system is said to have a degree of smoothness of order \textit{one}, that is $r=1.$ Hence the memristive MLC circuit can be considered to behave as a \textit{Filippov} system or a \textit{Filippov} flow capable of exhibiting \textit{sliding bifurcations}. 

Sliding bifurcations are Discontinuity Induced Bifurcations ( DIB's )
arising due to the interactions between the limit cycles of a Filippov system  with the boundary of a sliding region. Four types of sliding bifurcations have been identified by Feigin \cite{feigin94} and were subsequently analysed by di Bernado, Kowalczyk and others \cite{dib_ijbc01,kowal01,dib02,dib_ijbc03} for a general $n-$dimensional system. These four sliding bifurcations are \textit{crossing-sliding} bifurcations, \textit{grazing-sliding} bifurcations, \textit{switching-sliding} bifurcations and \textit{adding-sliding} bifurcations.

The memristive MLC circuit is found to admit three types of sliding bifurcations, namely \textit{crossing-sliding}, \textit{grazing-sliding} and \textit{switching sliding} bifurcations \cite{icha16}. Let the parameters be chosen as $a_{1,3} = -0.55$, $a_2 = -1.02$, $\beta= 0.95$, $f = 0.20$ and $\omega = 0.65$.  For these choice of parameters, the memristive MLC circuit undergoes repeated sliding bifurcations at the discontinuity boundaries $\Sigma_{1,2}$ and $\Sigma_{2,3}$, giving rise to a chaotic state as shown in Fig. \ref{fig:control1}. Here a(i) shows the time plot of the $x_1$ variable and a(ii) shows the phase portrait in the $(x_1-x_2)$ plane. In the subsequent section we will show that this chaotic behaviour exhibited by the memristive MLC circuit can be controlled using \textit{state feedback control} technique.

\section{Control of Chaos}
Control of chaos refers to purposeful manipulation of the chaotic behaviour of a nonlinear system to some desired or preferred dynamical state. As chaotic behaviour is considered undesired or harmful, a need was felt for suppression of chaos or at least reducing it as much as possible. For example, control of chaos is necessary in avoiding fatal voltage collapses in power grids, elimination of cardiac arrhythmias, guiding cellular neural networks to reach certain desirable pattern formations, etc. The earliest attempts at controlling chaos were focussed on eliminating the response of a chaotic system, which resulted in the destruction of the dynamics of the system itself. However it was Ott, Grebogi and Yorke \cite{ogy90} who showed that it would be beneficial to force the chaotic system to one of its infinite unstable periodic orbits (UPO) which are embedded in the chaotic attractor of the system without totally destroying the dynamics of the system. Following this many workers have developed newer techniques to control chaos and have applied them successfully on a variety of systems to realize different desired behaviours. Generally all the known methods of chaos control can be grouped into two categories, either
feedback control methods or non-feedback control algorithms.

\subsection{Feedback Controlling Algorithms}
Feedback control algorithms essentially make use of the intrinsic properties of chaotic systems to stabilize orbits which are already existing in the systems. The Adaptive Control Algorithm (ACA) developed by \cite{huberman90}  and applied by \cite{sinha90,rajasekar93}, the Ott-Grebogi-Yorke (OGY) Algorithm developed by \cite{ogy90} and applied by \cite{ding93,tel91,lai_tel93,ml96,singer91}, the Control Engineering Approach, developed by \cite{chen_dong92,chenG93} are all examples of these algorithms.  

\subsection{Non-feedback Methods}
The non-feedback methods refer to the use of some small perturbing external force, or noise, or a constant bias potential, or a weak modulating signal to some system parameter. The parametric control of chaos was demonstrated by, \cite{lima90,liu94,rajasekar93,wisen86a,wisen86b,braiman91,ml96}. The control of chaos by applying a constant weak biasing voltage was demonstrated by \cite{ml96} in the case of MLC oscillator and Duffing oscillator and by addition of noise was demonstrated in  a BVP oscillator by \cite{rajasekar93}. The other control algorithms are Entrainment or Open Loop Control method developed and applied by  \cite{jack90a,jack90b,jack91a,jack91b,jack91c}, the Oscillation Absorber Method developed by  \cite{kapit92,kapit93}.  

\subsection{Control of Chaos using State Feedback}
As the feedback and nonfeed back methods of chaos control have many drawbacks, 
a continuous time feedback control using small perturbations was proposed numerically by \cite{pyragas92}. This control scheme was provided a rigorous basis by  Chen and Dong and was demonstrated successfully in time continuous systems like Duffing Oscillator \cite{chen_dong93}, Chua's Circuit \cite{chenG93,hwang97} and so on. However the drawbacks of these methods are
\begin{enumerate}
\item
they can be applied only when the dynamical equations for the system are known \textit{a priori}
\item
the internal state variables are assumed to be available to construct control forces
\item
the controller structure, in some cases, is extremely complicated
\item
limited information may be available and the only measurable quantity of the system is its output
\item
Further, for nonsmooth systems, these conventional techniques, in particular addition of a second weak periodic excitation or the addition of a constant bias do not seem to enforce control of chaos
\end{enumerate}
Under such conditions a parallel state reconstruction by means of either a Kalman filter or Luenberger type observer must be used to implement control laws. For this purpose, the state space representation of the system and their transformations to either controller canonical form or observer canonaical form are derived, refer Appendix A. 

The state space representation refers to the modelling of dynamical systems in terms of state vectors and matrices so that the analyses of such systems are made conveniently in the time domain, using the basic knowledge of matrix algebra  \cite{kailath80,Ioannou96}. This representation is a well researched area in the field of control engineering \cite{ctchen70,vidya75,kailath80}. The main advantage of this approach is that it presents a uniform platform for representing time varying as well as time invariant systems, linear as well as piece-wise nonlinear systems. Further the vector fields for all the sub-spaces of the system take on a uniform form. Some of the methods of control that fall in this category are adaptive control \cite{yassen03}, observer based control \cite{liao98b}, sliding mode control \cite{yau04}, impulsive control \cite{sun03} and backstepping control \cite{yassen07}, linear switched state feedback method \cite{zhangj09}, twin-T notch filter method \cite{Iu11}, and backstepping method \cite{song11}.

In this section we outline the feedback method for control of chaos in a general dynamical system using state space models. Let us consider the observer canonical representation of a single input single output (SISO) nonlinear chaotic system in state space, refer Eq. (\ref{eqn:AppenA_ss2}) in Appendix A, 
\begin{eqnarray}
\dot{x}_o
 & = & \tilde{A}x_o + B^Tu,  \nonumber \\
y  		& = & C_o^Tx_o +D^Tu,
	\label{eqn:ss_openloop}
\end{eqnarray}
where $\tilde{A} \in \mathcal{R}^{n\times n}$, $B \in \mathcal{R}^{n \times r}$, $C \in \mathcal{R}^{n \times l}$ and $D \in \mathcal{R}^{l \times r}$ are matrices, $u$ is a $r$-dimensional vector denoting the control input and $y$ is a $l$-dimensional vector representing the output of the system. This system is often called as \textit{open-loop system} in control theory. 

Being in the observer canonical form, the system matrix $\tilde{A}$ is given as 
\begin{equation}
 \tilde{A} =  \left (	\begin{array}{ccccc}
				-\tilde{a}_1		&1	\;	\;	&0 	\;		\;	&\cdots		&0	\\
				-\tilde{a}_2		&0 	\;  \;  &1 	\;		\;	&\cdots		&0	\\
				\vdots		&\vdots	\;\;&\vdots	\;  \;	&\cdots		&\vdots	\\
				-\tilde{a}_{n-1}	&0	\;	\;	&0	\;		\;	&\cdots		&1	\\
				-\tilde{a}_n		&0	\;  \;	&0 	\;		\;	&\cdots		&0
				\end{array}
		\right ), 
\label{eqn:ss_Ao}
\end{equation}
where $\tilde{a}_i$'s are the coefficients of the characteristic polynomial $\{ |sI - \tilde{A}| \}$.  

If we want the states of the system to approach zero starting from any arbitrary state, then we should have to design a control input which would regulate the states of the system to the desired equilibrium conditions. To achieve this we assume a \textit{state feedback control law}
\begin{equation}
u = -\tilde{K} x_o,
	\label{eqn:feedback_cont}
\end{equation}
where $\tilde{K}$ is called the \textit{control gain vector} and can be designed using pole placement technique, familiar in control theory. 

\begin{figure*}
	\centering
	\resizebox{\textwidth}{!}
		{\includegraphics{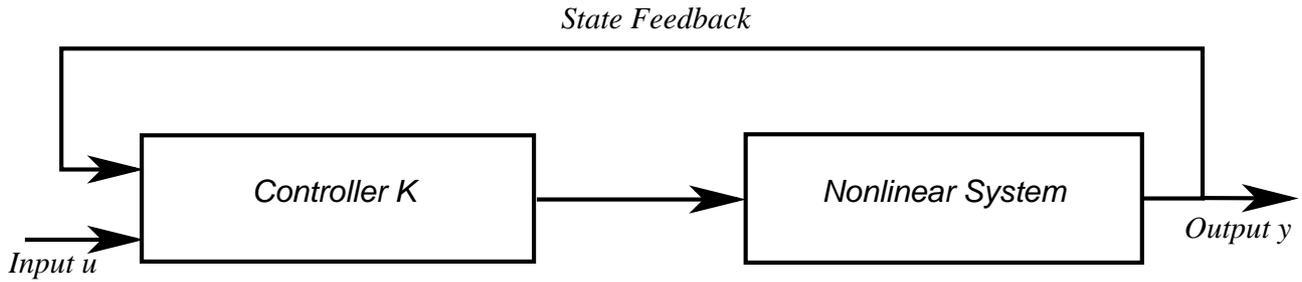}}		
	\caption[Block Diagram of State Feedback Control] {Block diagram illustrating the concept of state feedback control.}
	\label{fig:control_BD}
\end{figure*}

Substituting this control law, Eq. (\ref{eqn:feedback_cont}) in the state space representation of the open-loop system, Eq. (\ref{eqn:ss_openloop}), the system now becomes a \textit{closed-loop system} represented as
\begin{eqnarray}
\dot{x}_o & = & (\tilde{A}-B^T\tilde{K})x_o,  \nonumber \\
y  		& = & C_o^Tx_o +D^Tu, 
	\label{eqn:ss_closedloop}
\end{eqnarray}
where $B^T$ is the transpose of the vector B and the closed-loop system matrix is given as
\begin{equation}
 (\tilde{A}-B^T\tilde{K}) = 
 	 \left (	\begin{array}{lcccc}
		-(\tilde{a}_1-\tilde{k}_n)		&1	\;	\;	&0 	\;		\;	&\cdots		&0	\\
		-(\tilde{a}_2-\tilde{k}_{n-1})		&0 	\;  \;  &1 	\;	\;	&\cdots		&0	\\
	\;\;\;\;\;\vdots	&\vdots	\;	\;	&\vdots	\;  \;	&\cdots		&\vdots	\\
		-(\tilde{a}_{n-1}-\tilde{k}_2) &0	\;	\;	&0	\;		\;	&\cdots		&1	\\
		-(\tilde{a}_n-\tilde{k}_1)	&0	\;  \;	&0 	\;		\;	&\cdots		&0
				\end{array}
		\right ). 
\label{eqn:ss_A-BK}
\end{equation}
If the values of $\tilde{K}$ are so chosen that the eigen values of the matrix $(\tilde{A}-B^T\tilde{K})$ lie within the unit circle in the complex plane, then the system can be controlled to a desired stable equilibrium state. The problem of chaos control thus reduces to just determining a state feedback control gain vector $\tilde{K}$ such that the control law, Eq. (\ref{eqn:feedback_cont}), places the poles of the closed loop system, Eq. (\ref{eqn:ss_closedloop}), in the desired locations. An illustration of this concept is shown in the block diagram in Fig. \ref{fig:control_BD}.

A necessary and sufficient condition for successful pole placement is that the nonlinear system, that is, the pair of matrices $(\tilde{A},B)$, must be controllable. 

Let the characteristic polynomial $ \{sI -(\tilde{A}-B^T\tilde{K})\}$ of the closed-loop system, Eq. (\ref{eqn:ss_closedloop}), be given as 
\begin{equation}
s^n+(\tilde{a}_1-\tilde{k}_n)s^{n-1}+(\tilde{a}_2-\tilde{k}_{n-1})s^{n-2}+ \cdots +(\tilde{a}_n-\tilde{k}_1)=0.
		\label{eqn:charac_poly1}
\end{equation}
Let the characteristic equation of the desired control state of the system be
\begin{eqnarray}
(s-s_1)(s-s_2)(s-s_3)\cdots (s-s_n) & = &0, \nonumber \\
s^n+\alpha_1s^{n-1}+\alpha_2s^{n-2}+\cdots \alpha_{n-1}s+\alpha_n& = &0,
	\label{eqn:charac_poly2}
\end{eqnarray}
where $s_i$, $i=1,2,\cdots n$ are the desired poles to which the system should be guided and $\alpha_i$, $i=1,2,\cdots n$ are the coefficients of the desired characteristic equation.
By comparing Eqs. (\ref{eqn:charac_poly1}) and (\ref{eqn:charac_poly2}) we get
the elements of the transformed control gain vector $\tilde{K}$ as
\begin{equation}
\tilde{k}_n = \alpha_1-\tilde{a}_1,\;\;\nonumber
\tilde{k}_{n-1}  =  \alpha_2-\tilde{a}_2,\;\; \nonumber
\tilde{k}_{n-3}  =  \alpha_3-\tilde{a}_3,\;\; \cdots \nonumber 
\tilde{k}_1  =  \alpha_n-\tilde{a}_n.
	\label{eqn:K_values}
\end{equation}
\section{Control of Chaos in Memristive MLC Circuit}

In the earlier sections we have seen that the memristive MLC circuit is a piecewise-smooth dynamical system having two discontinuity boundaries causing the state space of the system to be split up into three sub-spaces. Consequently the memristive MLC circuit is represented by a set of smooth ODE's, refer Eqs. (\ref{eqn:smooth_odes}). Further we have seen that for the boundary equilibrium points given by  Eqs. (\ref{eqn:slid_points}), the memristive MLC circuit becomes a \textit{Filippov} system. 

Linearising the vector fields about the equilibrium points defined by Eqs. (\ref{eqn:slid_points}), the observer canonical form of the state space representation of the memristive MLC oscillator as a SISO system, refer  Eq. (\ref{eqn:AppenA_ss2}) in Appendix A, can be given as
\begin{eqnarray}
\dot{x}_o(t) & = & 
		\begin{cases}
		\tilde{A}_2 x_o +B^Tu & 	\text{if $x \in S_2$ }, \\
		\tilde{A}_{1,3} x_o+B^Tu & 		\text{if $x \in S_{1,3}$ },  \\
		\end{cases} \nonumber \\
y  		& = & C^Tx + D^Tu,
	\label{eqn:mmlc_ss1}
\end{eqnarray}
where the system matrices $\tilde{A}_i$'s are calculated for the above chosen parameters as 
\begin{equation}
 \tilde{A}_2(x)\;\; =  \left (	\begin{array}{ccccccc}
				\enspace 0.0700 &	&	&1.0	&  	&	&0.0	\\
				\enspace 0.0190	&  	&	&0.0 	&	&	&1.0	\\
				\enspace 0.0000	&	&	&0.0 	&	&	&0.0 	\\	
				\end{array}
		\right ),
\label{eqn:mmlc_A2}
\end{equation}
while 
\begin{equation}
 \tilde{A}_{1,3}(x) =  \left (	\begin{array}{ccccccc}
				-0.4000	&	&	&1.0	&	&	&0.0			\\
				-0.4275	&	&	&0.0 	&	&	&1.0			\\
				-0.0000	&	&	&0.0 	&	&	&0.0 			\\
				\end{array}
		\right ).
\label{eqn:mmlc_A13}
\end{equation}
Further the vectors $B^T$, $C^T$ and $D^T$ are chosen as
\begin{equation}
 B^T =  \left (	\begin{array}{ccc}
				0	&1		&0			\\
					\end{array}
		\right ),
\label{eqn:mmlc_B}
\end{equation}
\begin{equation}
 C^T =  \left (	\begin{array}{ccc}
				1	&0		&0			\\
					\end{array}
		\right ),
\label{eqn:mmlc_C}
\end{equation}
\begin{equation}
 D^T =  \left (	\begin{array}{ccc}
				0	&0		&0			\\
					\end{array}
		\right ).
\label{eqn:mmlc_D}
\end{equation}
We assume here that no disturbance is present in the system, that is, the vector $D^T$ is a null vector $D=0$.

\begin{figure*}[!t]
	\centering
	\resizebox{\textwidth}{!}
		{\includegraphics{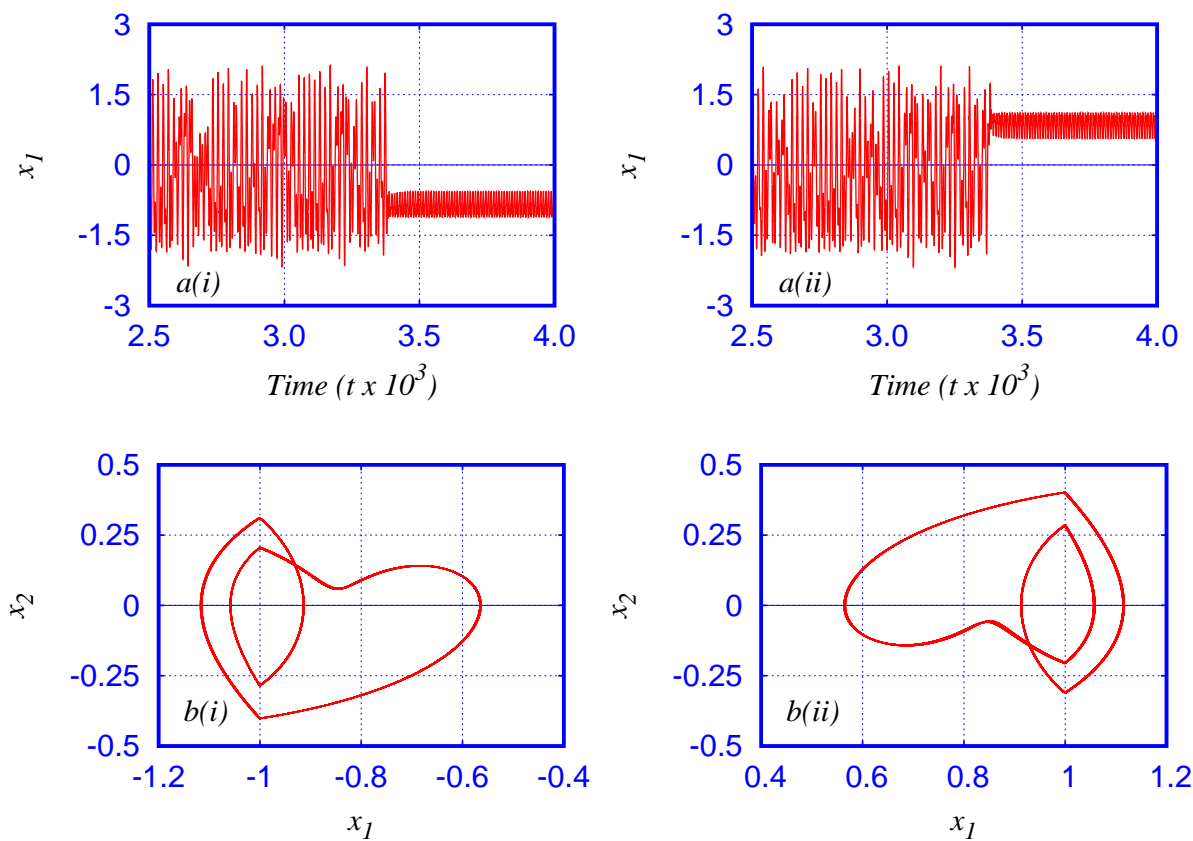}}
	\caption[Dynamics of the memristive MLC Circuit after application of state feedback control] {The periodic oscillations of the memristive MLC oscillator after the application of the state feedback control shown by a(i) \& a(ii) the time plots and b(i) \& b(ii) phase portraits in the $(x_1-x_2)$ plane. A change in the initial conditions form $(x_1 = -0.1,x_2 =-0.1,x_3 = -0.1)$ to $(x_1 = -0.2,x_2 =-0.2,x_3 = -0.2)$ results in the symmetric interchange of the time plots and attractors about the origin. The step size is assumed as $h = \frac{1}{1000}(2\pi/\omega)$, with $\omega = 0.65$ and $f = 0.20$.} 
	\label{fig:control2}
\end{figure*}
The peculiarity of this observer canonical representation, Eqs. (\ref{eqn:mmlc_ss1}), is that the transformations required become identical for all the three sub-regions of the phase space. This is particularly helpful in studying nonsmooth bifurcations of piecewise-smooth systems \cite{sontag98}.

The controllability matrices for the sub-spaces $S_{1,3}$ for the above mentioned parameters are given as
\begin{equation}
 P_{c_{1,3}} =  \left (	\begin{array}{ccccccc}
				0.0	&	&	&\enspace 1.00	&	&	&\enspace 0.5500		\\
				1.0	&	&	&\enspace 0.55 	&	&	&-0.6475			\\
				0.0	&	&	&-0.95 			&	&	&\enspace 0.3800 	\\	
				\end{array}
		\right ).
\label{eqn:mmlc_CO2}
\end{equation}
Similarly the controllability matrix for the sub-space $S_2$ is
\begin{equation}
 P_{c_2} =  \left (	\begin{array}{ccccccc}
				0.0	&	&	&\enspace 1.00	&	&	&\enspace 1.0200	\\
				1.0	&	&	&\enspace 1.02 	&	&	&\enspace 0.0904	\\
				0.0	&	&	&-0.95 			&	&	&-0.0665 			\\	
				\end{array}
		\right ).
\label{eqn:mmlc_CO13}
\end{equation}
As the controllability matrices in all the three sub-spaces have a full rank of $3$, we find that the matrices $(\tilde{A}_i,B)$ form controllable pairs. Hence the linearised parts of the memristive MLC circuit are controllable.
To achieve state feedback control, we assume a \textit{switched state feedback control law} \cite{zhangj09},
\begin{eqnarray}
u 	& = &
	\begin{cases}
	-\tilde{K}_2 x_o & 	\text{if $x \in S_2$ }, \\
	-\tilde{K}_{1,3} x_o &	\text{if $x \in S_{1,3}$ },  
	\end{cases}
	\label{eqn:feedback_cont1}
\end{eqnarray}
where $\tilde{K}_i$'s are the \textit{control gain vectors} in the three sub-regions of the phase space and are found using the procedure outlined in the previous section as
\begin{equation}
\tilde{K}_2\;\; = \left( \begin{array}{ccc}
					-0.2050  &	 0.8290 &	-1.2300
			  \end{array}
			  \right),
	\label{eqn:mmlc_k2}
\end{equation}
and
\begin{equation}
\tilde{K}_{1,3} = \left( \begin{array}{ccc}
					\;\;0.5040  &	1.4825 & \;\;2.0000
			  \end{array}
			  \right).
	\label{eqn:mmlc_k13}
\end{equation}
The \textit{closed loop system} for the memristive MLC circuit upon application of gain is
\begin{eqnarray}
\dot{x}_o(t) & = & 
		\begin{cases}
		(\tilde{A}_2\;\;-B^T\tilde{K}_2\;\;)x_o  & 	\text{if $x \in S_2$ }, \\
		(\tilde{A}_{1,3}-B^T\tilde{K}_{1,3}) x_o & 		\text{if $x \in S_{1,3}$ },  \\
		\end{cases} \nonumber \\
y  		& = & C_o^Tx + D^Tu.
	\label{eqn:mmlc_ss1k}
\end{eqnarray}
As the eigen values of the matrices $(\tilde{A}_i-B^T\tilde{K_i})$, $i = 1,2,3$ lie within the unit circle, the dynamics of the controlled closed system settles down to a non-chaotic equilibrium state. The chaotic attractor of the system before the application of the state feedback control and the controlled periodic state after the control has been applied are shown in Figs. \ref{fig:control2}.

The time series of the system which is chaotic before the application of control becomes periodic after the control is applied. This regulation of the chaotic time series to a periodic behaviour for the initial conditions $(x_1 = -0.1,x_2 =-0.1,x_3 = -0.1)$ is shown in Fig. \ref{fig:control2}a(i) while the periodic attractor in the $(x_1-x_2)$ phase plane in the asymptotic limit is shown in Fig. \ref{fig:control2}b(i).  However if the initial conditions are changed to $(x_1 = -0.2,x_2 =-0.2,x_3 = -0.2)$, we observe an inversion of the time series for the variable $x_1$ and the periodic attractor in $(x_1-x_2)$ phase space as are shown in the corresponding Figs. \ref{fig:control2}a(ii) and \ref{fig:control2}b(ii).

It is pertinent to state here that from Figs. \ref{fig:mmlc_limitcycle} and \ref{fig:control2} the memristive MLC system may possess multistability. This is because we see that in these two cases, a mere change in the initial conditions forces the system to exhibit different dynamics. If the system were to possess multistability, then we strongly believe that by tweaking the control gain vectors $K_1$ and $K_2$, it can be directed to take on any of the desired multistable states.

\section{Synchronization of Chaos}

The feasibility of synchronization of chaotic systems and the conditions to be satisfied for the same were first demonstrated by \cite{caroll90} by introducing the concept of \textit{Drive-Response} systems. Here a chaotic system is considered as the \textit{drive} system and a part of or subsystem of this drive system is considered as the \textit{response}. Under the right conditions ( the conditional Lyapunov exponents (CLEs) of the error dynamics being negative), the signals of the response part will converge to those of the drive system as time elapses. Ever since this ground breaking work, many researchers have proposed synchronization of chaos in different systems based on theoretical analysis and even experimental realizations. For example, this methodology has been successfully applied to synchronize chaos in Lorenz systems \cite{caroll90, caroll93b, vaidya92}, R\"{o}ssler systems \cite{caroll90}, the hysteretic circuits \cite{caroll91b}, Chua's circuits \cite{kocarev92a}, driven Chua's circuits \cite{kocarev92b}, Chua's and MLC circuits \cite{murali95, murali97}, ADVP oscillators \cite{kmurali93, kmurali94}, phase locked loops (PLL) \cite{endo91, sousa91}, etc. 

Further the possibility of applying this approach for secure communication has been demonstrated. The idea of \textit{chaotic masking and modulation} and \textit{chaotic switching} for secure communication of information signals based on Pecora and Caroll method of synchronization of chaos was demonstrated numerically by Cuomo and Oppenheim \cite{cuomo93a, cuomo93b, cuomo93c, cuomo94} and experimentally by Koracev \cite{kocarev92b}  using Chua's circuit as the chaos generator. Further the applicability of chaotic synchronization to digital secure transmission was demonstrated by \cite{cuomo93a} and experimentally by \cite{parlitz92, kmurali94}. The possibility of synchronization of  hyperchaotic systems and its applicability for communication purposes was proposed by \cite{peng96}. 
All these works make secure communications more practicable and with improved degree of security. 

Many alternative schemes of synchronization based on modifications of the drive-response concept have been proposed, such as the unidirectional coupling sche- me refer \cite{kmurali93,kmurali94}, function projective synchronization \cite{main99}, hybrid function projective synchronization \cite{chee03,xu01,xu02,grasmil09,grasmil07}, the arbitrary hybrid function projective synchronization \cite{huxu07b,luzhang08,huxu07a} etc. The synchronization of two canonical Chua's circuits using resistive unidirectional coupling has been studied by \cite{dvs05} and two unidirectional coupled SC-CNN based canonical Chua's circuits has been realised experimentally by \cite{swathi14}. The synchronization and propagation of a low frequency signal in a network of unidirectionally coupled Chua's circuits driven by a bi-harmonic external excitation has been studied by \cite{jothi13}. However all these methods have drawbacks such as,

\begin{enumerate}
\item
they do not give a systematic procedure for determining the response system and the drive signal. This means that most of the schemes are dependent on the drive system and could not be generalized to an arbitrary drive system.
\item
the dynamics of the drive system should be free of any disturbances.
\item
the conditional lyapunov exponents (CLE) should be negative. This condition restricts the signal to be transmitted to be a small perturbation to the state variables. As this requirement is not fulfilled by nonsmooth systems, such as in the case of a two coupled memristive MLC system, effecting synchronisation should necessary be obtained by other techniques only. 
\end{enumerate}

The concept of adaptive synchronization was applied by \cite{wu96, bernado96, liao98} and observer based approaches by \cite{morgul96, morgul97}  to overcome these difficulties of the drive-response concept.

\section{Observer Based Adaptive Synchronization of\\ Chaos}

Let us consider the state space representation of a single input single output (SISO) nonlinear system \cite{Ioannou96}, defined as in Eq. (\ref{eqn:AppenA_ss1}) in Appendix A,
\begin{eqnarray}
\dot{x} & = & \tilde{A}x + B^Tu,  \nonumber \\
y  		& = & C^Tx +D^Tu, 
	\label{eqn:ss_drive}
\end{eqnarray}
where $\tilde{A} \in \mathcal{R}^{n\times n}$, $B \in \mathcal{R}^{n \times r}$, $C \in \mathcal{R}^{n \times l}$ and $D \in \mathcal{R}^{l \times r}$ are matrices, $u$ is a $r$-dimensional vector denoting the control input and $y$ is a $l$-dimensional vector representing the output of the system. The control input can be given as
\begin{equation}
u = d +\theta^T f(x,y),
	\label{eqn:control}
\end{equation}
where $d \in R$ is a bounded disturbance, $\theta \in R^p $ is the constant parameter vector and $f(x,y)$ is a $p$-dimensional vector differential function.

When all the state variables of this system are unavailable for measurement, then according to control theory, the states of the system may be estimated by designing a parametric model of the original system. This parametric model is called an \textit{observer} and is considered as the response system. The concept of \textit{observer design} is a well established branch of control engineering and is widely used in the state feedback control of dynamical systems \cite{ctchen70,vidya75,kailath80}. In this method, once the drive system and its related observer are chosen, then under certain conditions, local or global synchronization between the drive and observer system is guaranteed \cite{morgul96}. 

Let us assume that the output $y(t)$ is the only variable that can be measured for the system Eq. (\ref{eqn:ss_drive}). Then an observer based on the available signal can be derived to estimate the state variables. This observer is known in the literature as the \textit{Luenberger Observer} \cite{Ioannou96} and is given as
\begin{eqnarray}
\dot{\hat{x}} & = & \tilde{A}\hat{x} + L^T(y-\hat{y})+ B^T\hat{u}, \nonumber \\					
\hat{y}		  & = & C^T \hat{x} +D^T\hat{u},
	\label{eqn:ss_response}
\end{eqnarray}
where $\hat{x}$ denotes the dynamic estimate of the state variable $x$, $L \in \mathcal{R}^n$ is a $n$-dimensional vector called as the \textit{observer gain vector}. It is essential that Eq. (\ref{eqn:ss_response})  is in observer canonical form, refer Eq. (\ref{eqn:AppenA_ss_Ao}) in Appendix A.

\begin{figure*}[!t]
	\centering
	\resizebox{1.9\columnwidth}{!}
		{\includegraphics{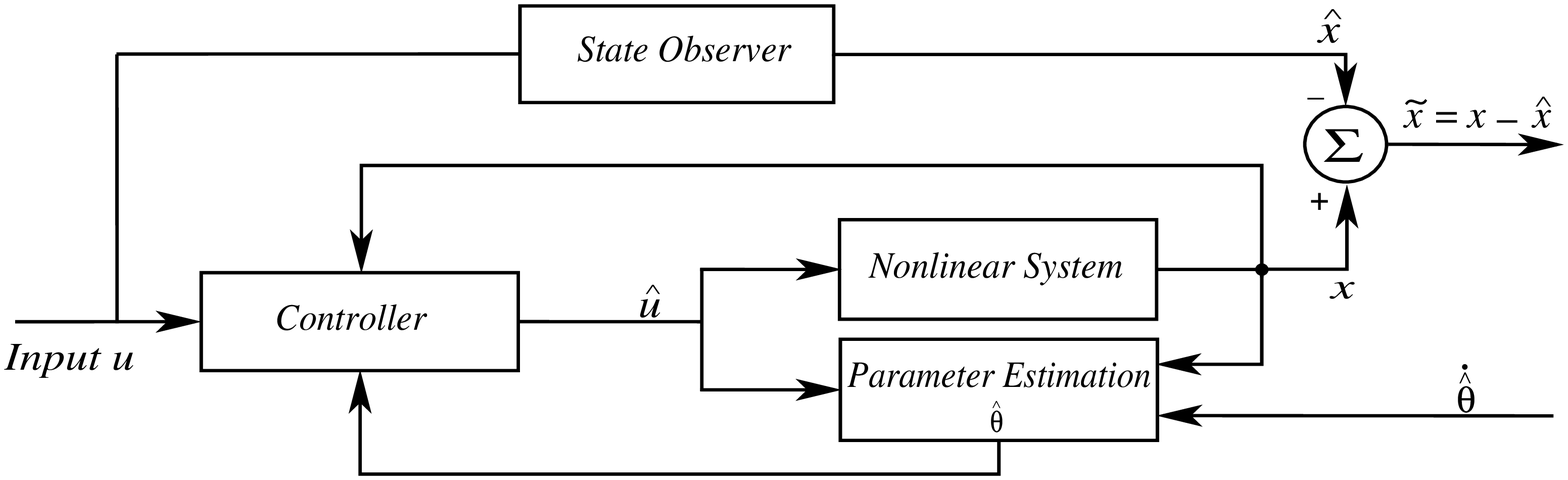}}
	\caption[Block Diagram Representation of Observer Based Adaptive synchronization] {Block diagrammatic representation of the observer based adaptive synchronization of nonlinear systems.} 
	\label{fig:sync_BD}
\end{figure*}

The control law can be derived as 
\begin{equation}
\hat{u} = \hat{d}+\hat{\theta}^Tf(x,y),
	\label{eqn:adap_control}
\end{equation}
where $\hat{d}$ and $\hat{\theta}$ are the estimates of the disturbances and the parameters of the system and are updated according to the adaptive algorithm \cite{liao2k} as
\begin{eqnarray}
\dot{\hat{d}} & = & (y-\hat{y}), \nonumber \\
\dot{\hat{\theta}} & = & f(x,y)(y-\hat{y}).
	\label{eqn:adap_algorithm}
\end{eqnarray}
The Luenberger observer Eqs. (\ref{eqn:ss_response}) has a feedback term that depends on the output observation error $\tilde{y}=y-\hat{y}$. Then the state observation error $\tilde{x}=x-\hat{x}$ satisfies the equation
\begin{eqnarray}
\dot{\tilde{x}} & = & (\tilde{A}-L^TC)\tilde{x} + B^T \left[(d-\hat{d})+(\theta^T - \hat{\theta}^T) f(x,y) \right], \nonumber \\
\tilde{x}(0) & = &  x_0-\hat{x}_0,
 	\label{eqn:error_dyn}
\end{eqnarray}
where we assume $X = (\tilde{A}-L^TC)$ as the augmented system matrix.
The implementation of this observer based adaptive synchronization of nonlinear systems is illustrated in Fig. \ref{fig:sync_BD}.\\

\subsection{Conditions for stability:}
According to control theory, the system represented by the Eq. (\ref{eqn:ss_response}) is stable in the sense of Lyapunov \cite{Ioannou96}, refer section A.1 of Appendix A, if any of the following conditions are satisfied:
\begin{enumerate}
\item
All eigen values of the augmented matrix $X = (\tilde{A}-L^TC)$, have negative real parts.
\item
For every positive definite matrix $Q$, (that is $Q = Q^T >0 $), the following Lyapunov matrix equation
\begin{equation}
X^TP+PX = -Q,
	\label{eqn:ss_lyp_eqn1}
\end{equation}
has a unique solution $P$ that is also positive definite.
\item
For any given matrix $C$, with the pair $(C,X)$ being observable, the equation
\begin{equation}
X^TP+PX = -C^TC,
	\label{eqn:ss_lyp_eqn2}
\end{equation}
has a unique solution $P$, that is also positive definite.
\end{enumerate}
\noindent If $(C^T,\tilde{A})$ is an observable pair, then we can choose the values of the gain vector $L$ such that the matrix $(\tilde{A}-L^TC)$ is stable. In fact, the eigen values of the matrix $(\tilde{A}-L^TC)$, and therefore the rate of convergence of $\tilde{x}(t)$ to zero can be arbitrarily chosen by designing the vector $L$  appropriately \cite{kailath80}.

The observer based response system given by Eq. (\ref{eqn:ss_response}) and associated with the control law given by Eq. (\ref{eqn:control}) and the adaptive algorithm given by Eq. (\ref{eqn:adap_algorithm}) will now globally and asymptotically synchronize with the drive system given by Eq. (\ref{eqn:ss_drive}), that is 
\begin{equation*}
\parallel \tilde{x}(t)\parallel = \parallel x(t)-\hat{x}(t)\parallel \rightarrow 0 \,\,\,\textrm{as}\,\,\, t \,\, \rightarrow \infty,
\end{equation*}
for all initial conditions.\\
\noindent Thus we find that the adaptive synchronization scheme is based on the following:
\begin{enumerate}
\item
the linear part of the system is observable, that is the pair $(C^T,\tilde{A})$ is observable,
\item
design of an suitable observer based on an adaptive law and
\item
formulation of a suitable control law.
\end{enumerate}

\section{Observer Based Adaptive synchronization of Chaos in Coupled Memristive MLC Oscillators}
\begin{figure}
	\centering
	
	\resizebox{\columnwidth}{!}
		{\includegraphics{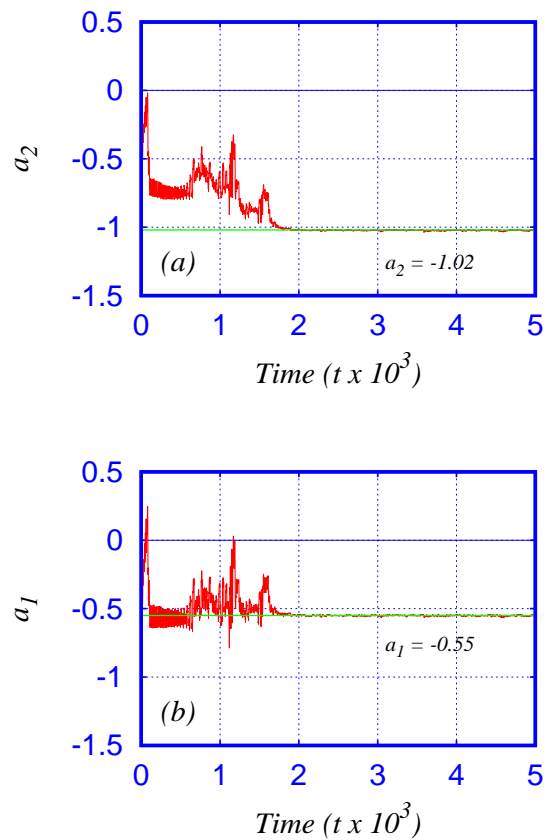}}
	\caption[Adaptive Observer Based Estimation of the Two Coupled Memristive MLC Circuit Parameters] { The estimation of (a) the parameter $a_2$ and (b) the parameter $a_1$ of response system of the two coupled Memristive MLC Circuit in the synchronized state using the adaptive observer scheme. It is to be noted that the asymptotic values of $a_2 = -1.02$ and $a_1 = -0.55$ are exactly equal to those of the drive system which were known apriori.}
	\label{fig:sync_parameters}
\end{figure}

In this section we report the synchronization of chaos via an observer based design, with appropriate control law and adaptive algorithm in a system of two coupled memristive MLC circuits. As in the case of control of chaos, we assume that under appropriate choice of the boundary equilibrium points, the memristive MLC circuit becomes a \textit{Filippov} system. Further we assume the same parameter values as were fixed for effecting control in a single memristive MLC circuit, namely $a_{1,3} = -0.55$, $a_2 = -1.02$ and $\beta= 0.95$, $f = 0.20$ and $\omega = 0.65$. Also we assume the observer canonical form of the state space representation of the memristive MLC circuit as given in Eq. (\ref{eqn:mmlc_ss1}), namely
\begin{eqnarray}
\dot{x}_o(t) & = & 
		\begin{cases}
		\tilde{A}_2 \enspace x_o +B^Tu & 	\text{if $x \in S_2$ }, \\
		\tilde{A}_{1,3} x_o+B^Tu & 		\text{if $x \in S_{1,3}$ },  \\
		\end{cases} \nonumber \\
y  		& = & C^Tx + D^Tu,
	\label{eqn:smmlc_ss1}
\end{eqnarray}
where the system matrices $\tilde{A}_i$'s and the vectors $B^T$, $C^T$ and $D^T$ are the same as are given in Eqs. (\ref{eqn:mmlc_A2} - \ref{eqn:mmlc_D}). Then the observability matrices for the sub-spaces $S_{1,3}$ are
\begin{equation}
 P_{o_{1,3}} =  \left (	\begin{array}{ccccccc}
				1.00	&	&	&0.00	&	&	&0.00			\\
				0.00	&	&	&1.00 	&	&	&0.00			\\
				0.00	&	&	&0.55	&	&	&1.00		 	\\	
				\end{array}
		\right ).
\label{eqn:smmlc_PO2}
\end{equation}
Similarly the observability matrix for the sub-space $S_2$ is
\begin{equation}
 P_{o_2} =  \left (	\begin{array}{ccccccc}
				1.00	&	&	&0.00	&	&	&0.00			\\
				0.00	&	&	&1.00 	&	&	&0.00			\\
				0.00	&	&	&1.02	&	&	&1.00 			\\	
				\end{array}
		\right ).
\label{eqn:smmlc_PO13}
\end{equation}
As these observability matrices in all the three sub-spaces have a full rank of $3$, we find that the matrices $(C^T,A_i)$ form an observable pair. Hence the linearised parts of the memristive MLC circuit are observable. Under this condition the Luenberger observer for the memristive MLC circuit can  be derived as
\begin{eqnarray}
\dot{\hat{x}}(t) & = & 
		\begin{cases}
		\tilde{A}_2 \enspace \hat{x} +L^T_2\enspace(y-\hat{y}) +B^T\hat{u} & 	\text{if $\hat{x} \in S_2$ }, \\
		\tilde{A}_{1,3} \hat{x}+ L^T_{1,3}(y-\hat{y}) +B^T\hat{u} &  \text{if $\hat{x} \in S_{1,3}$ },  \\		
		\end{cases} \nonumber \\
\hat{y}  		& = & C^T \hat{x} + D^T\hat{u},
	\label{eqn:mmlc_ss2}
\end{eqnarray}
where the control law is 
\begin{equation}
\hat{u} = \hat{\theta}^Tf(x,y),
	\label{eqn:sadap_control}
\end{equation}
with the vector $f(x,y)$ given as
\begin{eqnarray}
f_1(x,y) & = & y, \nonumber \\
f_2(x,y) & = & |y+1|-|y-1|,
	\label{eqn:sadap_fxy}
\end{eqnarray}
is the differential function and $\hat{\theta}$ are the estimates of the parameters of the system and are updated according to the adaptive algorithm
\begin{eqnarray}
\dot{\hat{\theta}}_1 & = & -(y-\hat{y})f_1(x,y), \nonumber \\
\dot{\hat{\theta}}_2 & = & -(y-\hat{y})f_2(x,y).
	\label{eqn:sadap_algorithm}
\end{eqnarray}
The state error $\tilde{x} = \dot{x}-\dot{\hat{x}}$ dynamics is represented by
\begin{eqnarray}
\dot{\tilde{x}} & = &
	\begin{cases}
(\tilde{A}_2\enspace -L^T_2\enspace C)\tilde{x} + B^T(\theta - \hat{\theta}^T)f(x,y),\\
(\tilde{A}_{1,3}-L^T_{1,3}C)\tilde{x} + B^T(\theta - \hat{\theta}^T)f(x,y). 
	\end{cases}	
	\label{eqn:smmlc_err}
\end{eqnarray}
The augmented matrices for the system can be defined as
\begin{equation}
X_i = (\tilde{A}_i\,-\,L^T_iC) \;\;\; \textrm{for} \;\;\;i=1,2,3.
	\label{eqn:mmlc_AM}
\end{equation}
For the choice of parameters of the system mentioned above, the observer gain vectors $L_i$ for each of the sub-spaces $S_i$ are chosen so as to have the augmented matrices $X_i$ to be exponentially stable.\\
For the sub-spaces $S_{1,3}$, the gain vectors are chosen as
\begin{equation}
L_{1,3} =  \left (	\begin{array}{ccc}
				0.8000	&	3.1000	&	-3.2870 		
				\end{array}
		\right )^T.
\label{eqn:mmlc_L13}
\end{equation}
Due to this choice of the observer gain vectors $L_i$, the augmented matrices $X_{1,3}$ in these sub-spaces $S_{1,3}$ will have \textit{poles} at 
$\{0.0000,\; -0.6000 \pm i 1.86748\}$.
Similarly, for the sub-space $S_2$, the gain vector is chosen as
\begin{equation}
L_2 =  \left (	\begin{array}{ccc}
				0.0000	&	15.2212	&	13.6508		
				\end{array}
		\right )^T.
	\label{eqn:mmlc_L2}
\end{equation}
This will cause the augmented matrix $X_2$ to have \textit{poles} at $\{-1.5788,\; 0.824398 \pm i 4.13832 \}$.\\
The Lyapunov equation for stability, Eq. (\ref{eqn:ss_lyp_eqn1}), may be written separately for the three sub-spaces as,
\begin{equation}
X_i^TP_i+P_iX_i = -Q \;\;\; \textrm{for} \;\;\;i=1,2,3,
	\label{eqn:ss_lyp_eqn3}
\end{equation}
where we assume the matrix $Q$ to be a $3$ - dimensional unit matrix. 
The solutions of the above Lyapunov equation for stability for the sub-spaces $S_{1,3}$ are positive definite matrices $P_{1,3}$ given as
\begin{equation}
 P_{1,3} =  \left (	\begin{array}{ccc}
				\;\;1.7778	&	-0.4220		&	-0.7308	\\
				-0.4220		&	 2.7528  	&	-0.1558	\\
				-0.7308		&	-0.1558		&	 0.6020 \\	
				\end{array}
		\right ).
\label{eqn:smmlc_P13}
\end{equation}
The matrix $P_2$ for the sub-space $S_2$ is given as 
\begin{equation}
 P_2\;\; =  \left (	\begin{array}{ccc}
				-0.0012		&	-0.0615		&	-0.0647	\\
				-0.0615		&	-3.0924 	&	-3.2555	\\
				-0.0647		&	-3.2555		&	-3.4268 \\	
				\end{array}
		\right ) \times 10^6,
\label{eqn:smmlc_P2}
\end{equation}
We find that the matrix $P_2$ for the sub-space $S_2$ is not a solution of the Lyapunov equation, Eq. (\ref{eqn:ss_lyp_eqn3}). Therefore the trajectories in this sub-space should be, as per Lyapunov theory, unstable. 
Hence the augmented matrix $X_2$ in region $S_2$ is also unstable. However the combined effect of the dynamics in the outer two sub-spaces $S_{1,3}$ represented by the augmented matrices $X_{1,3}$ and the positive definite matrices $P_{1,3}$ will impress upon the system as a whole to become asymptotically stable and exhibit a bounded behaviour asymptotically. Further as the conditions for the Lyapunov asymptotic stability, Eq. (\ref{eqn:ss_lyp_eqn3}) are satisfied by the system as a whole, we find that under the action of the control law, Eq. (\ref{eqn:sadap_control}) and the adaptive algorithm, Eq. (\ref{eqn:sadap_algorithm}), the estimated values of the unknown parameters of the observer system $\hat{a}_i$'s converge finally to the true values of the parameters $a_i$'s as time progresses. These are shown in Figs. \ref{fig:sync_parameters}, where we find in Fig. \ref{fig:sync_parameters}(a) the value of the parameter $\hat{a}_2$ converges to its true value of $-1.02$, while in Fig. \ref{fig:sync_parameters}(b) the value of the parameter $\hat{a}_1$ converges to its true value of $-0.55$.

\begin{figure}[!t]
	\centering
	\resizebox{\columnwidth}{!}
		{\includegraphics{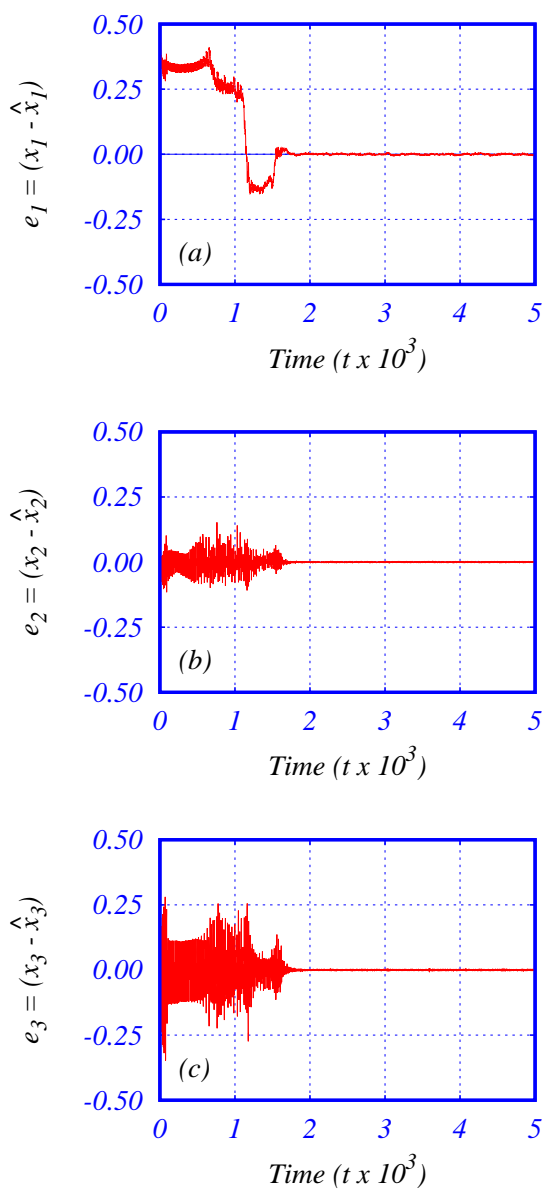}}
	\caption[Error Dynamics of the Two Coupled Memristive MLC Circuit] {The convergences of the errors in the variables, $e_1 = x_1-x_1'$, $e_2 = x_2-x_2'$ and $e_3 = x_3-x_3'$ of the two coupled Memristive MLC Circuit in synchronized state under adaptive observer scheme are shown in (a), (b) and (c) respectively.}
	\label{fig:sync_error}
\end{figure}

\begin{figure}[!t]
	\centering
	\resizebox{\columnwidth}{!}
		{\includegraphics{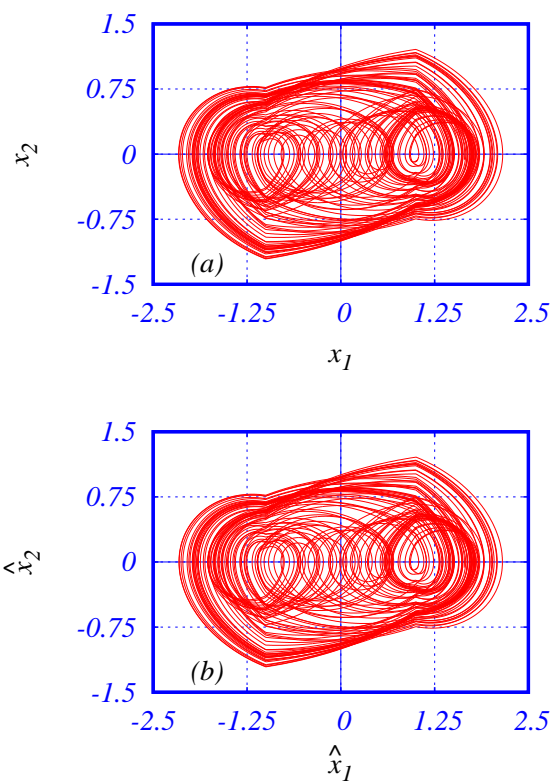}}
	\caption[Phase portraits of the drive and response Memristive MLC Circuits] {The phase portraits (a) in the $(x_1-x_2)$ plane and (b) in the $(\hat{x}_1-\hat{x}_2)$ plane showing identical chaos respectively.}
	\label{fig:sync_chaos}
\end{figure}

Mathematically we have the error between the drive and the response, converging to zero for all initial values, as time progresses, that is  
\begin{equation*}
\parallel \tilde{x}(t)\parallel = \parallel x(t)-\hat{x}(t)\parallel \rightarrow 0 \,\,\,\textrm{as}\,\,\, t \,\, \rightarrow \infty.
\end{equation*}
The convergence of the error dynamics $\tilde{x}$ to zero is shown in Fig. \ref{fig:sync_error}. Here the convergence of the errors $\tilde{x}_1$,  $\tilde{x}_2$ and  $\tilde{x}_3$ are shown in plots (a), (b) and (c) of Fig. \ref{fig:sync_error} respectively.
These convergences of the parameters to their true values and that of the error dynamics to zero, cause the observer system dynamics to converge to the original system dynamics as time elapses. This means that the response system dynamics evolves as time proceeds to that of the drive system dynamics. Hence if the drive system is in a chaotic state, then the response system should also exhibit identical chaotic state. This is shown in Fig. \ref{fig:sync_chaos}.

\begin{figure}[!t]
	\centering
	\resizebox{\columnwidth}{!}
		{\includegraphics{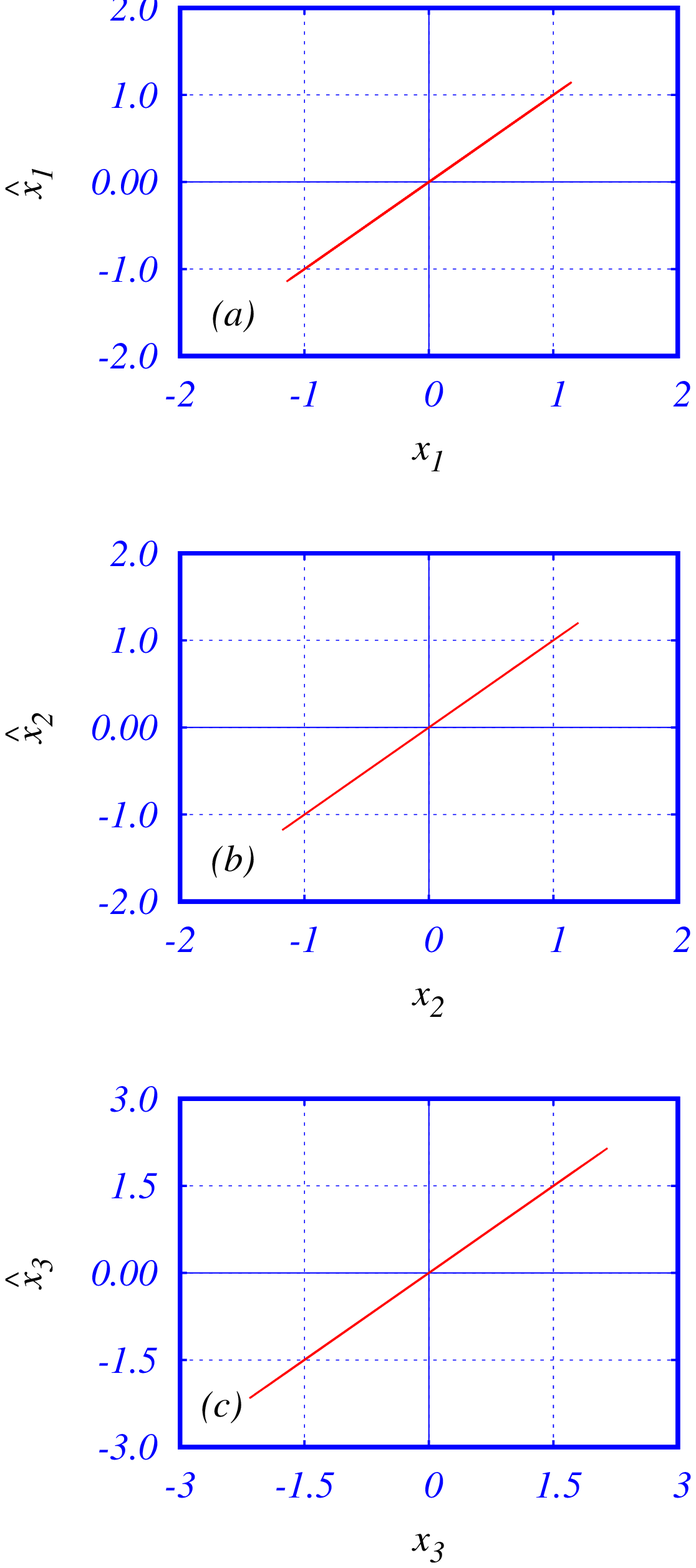}}
	\caption[Complete synchronization of the Two Coupled Memristive MLC Circuit] {The complete synchronization of the two coupled Memristive MLC circuit under the adaptive observer scheme, shown in (a) the $(x_1-x_1')$ plane, (b) the $(x_2-x_2')$ plane and (c) the $(x_3-x_3')$ plane.}
	\label{fig:sync_variable}
\end{figure}

Had the drive system been in a periodic state, then one would expect the response system also to take on asymptotically the periodic state by virtue of the adaptive synchronization. As both the drive and the response systems exhibit identical behaviour, they are said to be in \textit{complete synchronization} (CS) with each other. This is shown by the diagonal lines for the variables in the $(x_1-x'_1)$, $(x_2-x'_2)$ and $(x_3-x'_3)$  phase planes in plots (a), (b) and (c) respectively in Fig. \ref{fig:sync_variable}. 

For effecting this, it is essential that the gain vectors $L_i$ for all the sub-spaces $S_i$'s are properly chosen. Due to the differences in the gain vectors in the three sub-regions of the phase space, this observer based adaptive synchronization is also referred to in literature as \textit{switched state feedback} method of adaptive synchronization \cite{zhangj09}.

\section{Conclusion}
In this work, we have studied the control of chaos in an individual memristive MLC circuit as well as the synchronisation behaviour in a system of two coupled memristive MLC circuits using \textit{state feedback control} and \textit{observer based adaptive control} techniques respectively. To realize these objectives, we have considered the memristive MLC circuit as a \emph{Filippov} system, a non-smooth system having the order of discontinuity \emph{one} and have derived the discontinuity mapping corrections such as (ZDM and PDM). Further we have derived the canonical state space representations for memristive MLC circuit. Also the stability theory of Lyapunov and pole-placement methods, concepts which are very much familiar in control theory, were applied. 

We wish to state here that we have derived analytical conditions for effecting control and adaptive synchronization using state feedback and implemented the results using numerical simulations. The fact that the results of simulations agree with the predictions of the analytical conditions point to the validity of our derivations.

From a different point of view, it has been shown by many researchers, that in general any two coupled systems, be they smooth or discontinuous, can be directed towards amplitude death or oscillation death, irrespective of their being in periodic, chaotic, hyper-chaotic or time-delay systems, by the application of proper feedback coupling, for example see \cite{resmi2011}. The same can be applied to the two coupled system under study, by calculating proper observer gain vectors and choosing proper initial conditions and parametric values. However we have not proceeded along these lines because it falls beyond the realm of this present work. We hope to pursue this possibility in future studies.

The phenomenon of control of chaos may be further studied to understand and effectively prevent the incidence of nonlinear catastrophic phenomena such as {\textit{blackouts}} in transmission lines and power grids, cardiac arrythmias, etc. The synchronisation of chaos which we have demonstrated using observer based adaptive scheme in memristive MLC circuits can be used to effect digital modulation schemes for secure communication. For example, the modulation characteristics of the memristor can be used to implement Amplitude Shift Keying ASK, a key technique in Digital Signal Processing and transmission of Digitized Information. Also the switching characteristics of the memristor can be utilised to implement digital protocols for secure transmission of data.

\appendix

\section{Space Representations of Dynamical Systems}
\noindent The state space representation refers to the modelling of dynamical systems in terms of state vectors and matrices so that the analyses of such systems are made conveniently in the time domain, using the basic knowledge of matrix algebra. The main advantage of this approach is that it presents a uniform platform for representing time varying as well as time invariant systems, linear as well as piece-wise nonlinear systems. The theoretical details presented here are essentially from the available literature on control systems \cite{kailath80,Ioannou96}. 

The generic state space representation of a $n^{th}$- order dynamical system is given as
\begin{eqnarray}
\dot{x} & = & Ax + Bu,  \nonumber \\
y  		& = & C^Tx +Du, 
	\label{eqn:AppenA_ss1}
\end{eqnarray}
where $x$ is an $n$-dimensional vector representing the state variables, $A \in \mathcal{R}^{n\times n}$, $B \in \mathcal{R}^{n \times r}$, $C \in \mathcal{R}^{n \times l}$ and $D \in \mathcal{R}^{l \times r}$ are matrices, $u$ is a $r$-dimensional vector denoting the control input and $y$ is a $l$-dimensional vector representing the output of the system.

The first of Eq. (\ref{eqn:AppenA_ss1}) is referred to as the \textit{state equation} while the second is referred to as the \textit{output equation}. The solution of the state equation is given by
\begin{equation}
x(t)=e^{A(t-t_0)}x(t_0) +\int_{t_0}^te^{A(t-\tau)}Bu(\tau)d\tau,
	\label{eqn:AppenA_ss_soln}
\end{equation}
where $e^{At}\equiv \Phi(t)$ is the \textit{state transition matrix} and $x(t_0)$ is the initial state of the system.
\begin{enumerate}
\item 
\textbf{Open-Loop System:} 
\noindent If the output of the system is neither fedback to the input nor is it used to modulate the behaviour of the system, then the system is called as a \textit{open-loop system}.\\

\item
\textbf{Closed-Loop System:}
\noindent If the output of the system is used to modulate the system and manipulate the control action on the system through some suitable feed-back mechanism, then the system is known as \textit{closed-loop system}.\\

\item
\textbf{Exponential Stability:}
\noindent An equilibrium state $x_e$ is said to be exponentially stable, if there exists a constant $\alpha > 0$, and for every small constant $\epsilon > 0$ there exists a small neighbourhood $|x_0-x_e|<\delta(\epsilon)$ such that 
\begin{equation}
|x(t;t_0,x_0)-x_e|\leq \epsilon e^{-\alpha(t-t_0)}.
	\label{eqn:AppenA_es}
\end{equation}
Here $\alpha$ is called as the rate of convergence.\\

\item
\textbf{Asymptotic Stability:} 
\noindent  An equilibrium state $x_e$ is said to be asymptotically stable in the sense of Lyapunov \cite{Ioannou96} if any of the following conditions are satisfied:
\begin{enumerate}
\item
All eigen values of the matrix $A$ have negative real parts.
\item
For every positive definite matrix $Q$, (that is $Q = Q^T >0 $), the following Lyapunov matrix equation
\begin{equation}
A^TP+PA = -Q,
	\label{eqn:AppenA_as1}
\end{equation}
has a unique solution $P$ that is also positive definite.
\item
For any given matrix $C$, with the pair $(C^T,A)$ being observable, the equation
\begin{equation}
A^TP+PA = -C^TC,
	\label{eqn:AppenA_as2}
\end{equation}
has a unique solution $P$, that is also positive definite.\\
\end{enumerate}

\item
\textbf{Observability:}
\noindent It refers to the determination of the state of a system by observing or measuring its output. Mathematically it is determined by finding the rank of the \textit{observability matrix}
\begin{equation}
P_o = \left( \begin{array}{c}
				C^T  \\
				C^TA  \\
				C^TA^2\\
				.  	\\
				.  	\\
				. 	\\
				C^TA^{n-1} \\
			\end{array}
		\right).
		\label{eqn:AppenA_Po}
\end{equation}
The observability matrix is of dimension $n \times nl$. If this observability matrix has a full rank, equal to $n$, then the dynamical system or the pair $(C^T,A)$ is said to be observable. However if $P_o$ is a $n \times n$ square matrix, then the system is observable if $P_o$ is non-singular.\\

\item
\textbf{Detectability:}
\noindent A dynamical system may not be completely observable. However if the \textit{unobservable} parts of the system become asymptotically stable under the action of some \textit{control law}, then the system is called as \textit{detectable} \cite{kailath80}.\\

\item
\textbf{Controllability:} 
\noindent It refers to the transferring of a system from any given initial $x(t_0)$ to any given desired final state $x(t_f)$ over a finite interval of time $(t_f-t_0)$. Mathematically it is determined by the rank of the \textit{controllability matrix}
\begin{equation}
P_c = \left( \begin{array}{c}
				B  \nonumber \\
				AB \nonumber \\
				A^2B \nonumber \\
				. \nonumber \\
				. \nonumber \\
				. \nonumber \\
				A^{n-1}B \nonumber \\
			\end{array}
		\right).
		\label{eqn:AppenA_Pc}
\end{equation}
The controllability matrix is of dimension $n \times nr$. If this controllability matrix has a full rank, equal to $n$, then the dynamical system or the pair $(A,B)$ is said to be controllable. However if $P_c$ is a $n \times n$ square matrix, then the system is controllable if $P_c$ is non-singular.\\

\item
\textbf{Stabilizability:} 
\noindent A dynamical system may not be completely controllable. However if the \textit{uncontrollable} parts of the system become asymptotically stable under the action of some \textit{control law}, then the system is called as \textit{stabilizable} \cite{kailath80}.
\end{enumerate}

\section*{Forms of State Space Representations}
\noindent For any given dynamical system, there are essentially an infinite number of possible state space models that give identical input/output dynamics. However it is often desirable to have certain standardized state space model structures called as the canonical forms or canonical state space representations. Using similarity transformations it is possible to convert the state space model from one canonical form to another \cite{kailath80}. Two of the most important canonical forms in control theory are the \textit{observer canonical form} and the \textit{controller canonical form}.

\subsection*{Observer Canonical Form}
\noindent Let us consider the coordinate transformation ${x_o} = W x$, where $W = T P_o$ is a transformation matrix, $P_o$ the observability matrix and the matrix $T$ is constructed using the coefficients of the characteristic polynomial of the state matrix $A$.
\begin{equation}
 T =  \left (	\begin{array}{ccccc}
			1				&0				&0		&\cdots		&0		\\
			\tilde{a}_1		&1    			&0		&\cdots		&0		\\
			\tilde{a}_2		&\tilde{a}_1 	&1		&\cdots		&0		\\
			\vdots			&\vdots			&\vdots	&\cdots		&\vdots	\\
			\tilde{a}_{n-1}	&\tilde{a}_{n-2}&\cdots	&\cdots		&1	
				\end{array}
		\right ). 
	\label{eqn:AppenA_ss_T}
\end{equation}
The characteristic polynomial $\{|s I - A|\}$ itself is given as
\begin{equation}
p(s) = s^n+\tilde{a}_1s^{(n-1)}+\tilde{a}_2s^{(n-2)}+...............\tilde{a}_{(n-1)}s+\tilde{a}_n.
	\label{eqn:AppenA_charac_poly}
\end{equation}
Using the inverse coordinate transformation $x = W^{-1}x_o$, Eq. (\ref{eqn:AppenA_ss1}) can be transformed to the observer canonical form as
\begin{eqnarray}
\dot{x}_o & = & \tilde{A}_ox_o + B^Tu,  \nonumber \\
y  		& = & C^T_ox_o +D^Tu, 
	\label{eqn:AppenA_ss2}
\end{eqnarray}
where the state matrix $A_o$ is obtained by the similarity transformation $A_o = W A W^{-1}$ and is given as
\begin{equation}
 \tilde{A}_o =  \left (	\begin{array}{ccccc}
-\tilde{a}_1		&1	\;	\;		&0 	\;		\;	&\cdots		&0		\\
-\tilde{a}_2		&0 	\;  \;  	&1 	\;		\;	&\cdots		&0		\\
\vdots				&\vdots	\;	\;	&\vdots	\;  \;	&\cdots		&\vdots	\\
-\tilde{a}_{n-1}	&0	\;	\;		&0	\;		\;	&\cdots		&1		\\
-\tilde{a}_n		&0	\;  \;		&0 	\;		\;	&\cdots		&0
				\end{array}
		\right ), 
\label{eqn:AppenA_ss_Ao}
\end{equation}
and 
\begin{equation}
C_o^T = C^T W^{-1}.
	\label{eqn:AppenA_Co}
\end{equation}

\subsection*{Controller Canonical Form}
\noindent  Let us in this case, consider an alternate coordinate transformation ${x_c} = M x$, where $M = T P_c$ is a transformation matrix, $P_c$ the controllability matrix and the matrix $T$ is constructed using the coefficients of the characteristic polynomial of the state matrix $A$ and is as given in Eq. (\ref{eqn:AppenA_ss_T}).

Using the inverse coordinate transformation $x = M^{-1}x_c$, Eq. (\ref{eqn:AppenA_ss1}) can now be transformed alternatively into the controller canonical form as
\begin{eqnarray}
\dot{x}_c & = & \tilde{A}_cx_c + B^Tu,  \nonumber \\
y  		& = & C^T_cx_c +D^Tu, 
	\label{eqn:AppenA_ss3}
\end{eqnarray}
where the state matrix $A_c$ is obtained by the similarity transformation $A_c = M A M^{-1}$ and is given as
\begin{equation}
 \tilde{A}_c =  \left (	\begin{array}{ccccc}
				0				&1	\;	\;			&0 	\;		\;		&\cdots				&0	\\
0				&0 	\;  \;  		&1 	\;		\;		&\cdots				&0	\\
\vdots			&\vdots	\;	\;		&\vdots	\;  \;		&\cdots				&\vdots	\\
0				&0	\;	\;			&0	\;		\;		&\cdots				&1	\\
-\tilde{a}_1	&-\tilde{a}_2\; \;	&-\tilde{a}_3\;	\;	&-\tilde{a}_{n-1}	&-\tilde{a}_n
				\end{array}
		\right ), 
\label{eqn:AppenA_ss_Ac}
\end{equation}
and 
\begin{equation}
C_c^T = C^T M^{-1}
	\label{eqn:AppenA_Cc}.
\end{equation}

\section*{Acknowledgement}
This work has been supported by a DST-SERB Distinguished Fellowship to M.L.

\bibliography{Bibliography}
\end{document}